\documentclass{aa}
\usepackage{amsmath}
\usepackage{amsfonts,color}
\usepackage{tensor}
\usepackage{amssymb,float}
\usepackage[utf8]{inputenc}
\usepackage{xcolor}
\usepackage[unicode=true,pdfusetitle,
 bookmarks=true,bookmarksnumbered=true,bookmarksopen=true,bookmarksopenlevel=3,
 breaklinks=false,pdfborder={0 0 1},backref=false,colorlinks=true]
 {hyperref}
\usepackage{enumitem}
\usepackage{graphicx}
\usepackage{appendix}
\usepackage{hyperref}
\usepackage{lipsum}
\usepackage{mathtools}
\usepackage{soul}
\usepackage[varg]{txfonts}

\makeatletter
\long\def\dddddot#1{%
  {\mathop {#1}\limits ^{\vbox to-1.4\ex@ {\kern -\tw@ \ex@ \hbox {\normalfont .....}\vss }}}%
}
\long\def\multidots#1#2{%
  \count@=0
  {{\mathop {#2}\limits ^{\vbox to-1.4\ex@ {\kern -\tw@ \ex@ \hbox {\normalfont %
  \loop%
  \ifnum#1>\count@%
  .%
  \advance\count@ by1%
  \repeat%
  }\vss }}}}%
}
\makeatother

\begin{document}

\title{On Negative Mass Cosmology in General Relativity}

 \author{Sebastián Nájera
          \and
          Aldo Gamboa\and Alejandro Aguilar-Nieto \and Celia Escamilla-Rivera
          }

   \institute{Instituto de Ciencias Nucleares, Universidad Nacional Aut\'onoma de M\'exico, 
Circuito Exterior C.U., A.P. 70-543, M\'exico D.F. 04510, M\'exico.
             \email{najera.sebastian@ciencias.unam.mx} }

\abstract{
In this Letter we present strong arguments in favour of thoroughly revising the negative mass cosmology (NMC), which has been proposed as a simple alternative explanation of dark energy and dark matter effects, within the framework of General Relativity. We show that there are various physical predictions of this model which require fine-tuning in order to make them compatible with current cosmological surveys. In this way, the original motivation of the NMC model becomes obscured due to the imposition of fine-tuned unknown variables. We conclude that a more rigorous theoretical treatment is needed in order to make the NMC a viable cosmological model.}


\maketitle


\textit{Introduction}.-
A wide variety of observations currently favour the standard model of cosmology $\Lambda$CDM, which one of its intrinsic theoretical problems is the nature of the \textit{dark sector} of the Universe. Along the years, several proposals that span from theoretical and observational points of view have treated successfully this issue, too many that to give fair references in this introduction would be an extensive task. In this Letter we will focus our discussion in one particular proposal:
a physical model involving continuously created negative masses in cosmology \citep{farnes}, which we will denote as NMC. This proposal is presented as
an alternative explanation to the dark sector of the Universe in a unified framework based on a Robertson-Walker (RW) geometry.

The existence of negative masses is not a new idea \citep{1957RvMP...29..423B, 1989GReGr..21.1143B}.  Moreover, it naturally originates problems with theorems of energy conditions (see e.g.~\citealt{Schon:1979rg, witten}). Nevertheless, the existence of physical phenomena which violate some of these theorems, led one to question their validity (see e.g.~\citealt{Barcelo:2002bv} for a review on violations of various energy conditions). Thus, despite its weirdness, the proposal of negative masses has been used in several physical scenarios. For example, it has been used to replace dark energy by a negative matter action \citep{Petit:2014ura}, and, in a Schwarzschild space--time, it has been found that there is no direct connection between the regions settled by the negative and positive masses \citep{Sch1}.

Both positive and negative masses can coexist in the Universe under the framework of GR, which assumes the validity of the weak equivalence principle. This assumption naturally leads to the conclusion that positive mass particles attract all massive particles, and  negative mass particles repel all massive particles  \citep{1957RvMP...29..423B}. However, the coexistence of negative and positive masses triggers a peculiar phenomenon known as \textit{runaway motion}, where two particles of equal and opposite mass produce a constant acceleration of the system towards the positive mass particle. This effect could have serious implications in structure formation and the expansion of the Universe. Therefore, we believe that a serious and rigorous analysis of this effect is pertinent and should be performed.

Before starting this analysis, we should mention that the NMC model presented by \cite{farnes} has received some strong critics, e.g. \cite{Socas-Navarro:2019pps} identified several inconsistencies of this model with cosmological observations, such as incorrect galactic halo masses, runaway galactic motions or a different rate of cosmological structure formation. Also, \cite{Stepanian:2019ncv} made a heuristic analysis of the NMC model considering energy conditions and observations from the early Universe, resulting in a mismatch. However, our analysis discussed in this Letter goes beyond these ideas.\\

\textit{Negative Mass Cosmology}.-
\label{sec:neg mass cosmology}
In the NMC model the total density parameter can be written as the contribution of each component of the Universe as
\begin{equation} \label{eq:omegas}
    \Omega=\Omega_{M+}+\Omega_{M-}+\Omega_{\Lambda},
\end{equation}
where $\Omega_{M+}$, $\Omega_{M-}$ and $\Omega_{\Lambda}$ are the critical density parameters of positive mass, negative mass and cosmological constant, respectively, and $\Omega_k=1-\Omega$ is the curvature critical density parameter.

In this Universe we can have three different scenarios: (i) a positive-mass dominated Universe, $|\Omega_{M+}|>|\Omega_{M-}|$, (ii) a massless cosmology, $|\Omega_{M+}|=|\Omega_{M-}|$, in which there is an equal number of positive and negative particles, and (iii) a negative-mass dominated Universe, $|\Omega_{M+}|<|\Omega_{M-}|$, with special attention given to the latter scenario  as  it deals with a possible cosmological constant effect produced by the continuously created negative masses and thus there is no need for a cosmological constant ($\Lambda=0$).

As it is standard in a RW space-time, the Friedmann equation is given by\footnote{Throughout this Letter we will use geometrical units.}
\begin{equation} \label{friedmanneq1}
     H^2 = \frac{\kappa}{3} ( \rho_+ + \rho_-) + \frac{ \Lambda}{3} - \frac{k}{a^2},
\end{equation}
where $H$ is the Hubble parameter, $\kappa = 8\pi$, $a$ is the scale factor, $k$ is the curvature parameter, $\rho_+$ is the positive mass density and $\rho_-$ is the negative mass density associated with a pressure $p_-$ through an equation of state $w_- = p_-/\rho_-$ for the negative masses. Throughout this Letter we leave $w_-$ general, unless specifically stated. According to the NMC model presented by \cite{farnes}, in the case where $\Lambda \neq 0$, there is a degeneracy between $\Omega_{M-}$ and $\Omega_{\Lambda}$, which is given by $\Omega_{\textrm{degen}}=\Omega_{M-}+\Omega_{\Lambda}$, and it is claimed  that in a conventional $\Lambda$CDM cosmology we are measuring this degeneracy. Thus, by taking $\Omega_{M-}$ to be zero, we are falsely inferring a $\Lambda$ instead of a negative mass density parameter.

The degeneracy previously shown can be discriminated by the equation of state $w_-$ of the negative masses, which yields $w_-=0$ for non-relativistic matter. To resolve this degeneracy, the NMC model considers that negative matter is constantly created by adding a \textit{matter creation term} to the Einstein field equations with $\Lambda$. This will result in an effective equation of state for the negative mass fluid in which $w_{\textrm{eff}}\neq 0$. 
By adding the creation term, the Einstein field equations,
\begin{equation}\label{eqTeff}
    G_{ab} + \Lambda g_{ab} = \kappa T_{ab},
\end{equation}
where $G_{ab}$ is the Einstein's tensor, $g_{ab}$ is the Robertson-Walker metric, will be modified so that the energy-momentum tensor $T_{ab}$ is 
\begin{equation} \label{eq:energymom nmc}
    T_{ab} = (\rho + p + P_c) u_au_b+\left(p+P_c\right)g_{ab},
\end{equation}
with $\rho$ the total mass density, $p$ the total pressure, and
\begin{equation}\label{eq:defPc}
    P_c \coloneqq -\frac{\Gamma}{3H}(\rho_- + p_-),
\end{equation}
an effective pressure caused by the creation of negative mass particles, where $\Gamma$ is the creation rate which, in principle, could depend on space and/or time. The associated continuity equation for the negative masses is
\begin{equation}\label{eq:continuity}
    \dot \rho_- + 3 H (\rho_- + p_-) = \Gamma (\rho_- + p_-).
\end{equation} 
See e.g.~\cite{Pan_2016} for additional details of the background creation of particles in cosmology. We note that the NMC model with $\Gamma = 0$ refers to the conventional negative mass cosmology within GR without creation of negative mass particles. 

The effective equation of state parameter for a Universe in which matter is constantly created resembles that of a cosmological constant when $\Gamma=3H$. Therefore, this NMC \textit{toy} model has what appears to be an effective cosmological constant $\Lambda_- \coloneqq 8\pi G \rho_- < 0$. If we take $\rho_+ = \Lambda = 0$ in this toy model, we can obtain the evolution of the scale factor in a negative mass-dominated cosmology from Eq.~\eqref{friedmanneq1}. The solution corresponds to an anti-de Sitter (AdS) Universe that undergoes a cycle of expansion and contraction with a timescale of $\sqrt{- 3\pi^2/\Lambda_-}$. This cyclic cosmology does not match with our current observations, however \cite{farnes} argued that there are two possible explanations for this: (i) either the Universe is so large such that the local geometry appears to be flat, or (ii) the presence of negative matter and its creation would modify the CMB anisotropies. Both arguments would make the NMC toy model compatible with early time observations.
A few examples that employ this NMC toy model are included in dark matter halos formation simulations, structure formation simulations and also, in galaxy rotation curves, by means of cosmological N-body simulations. Outside from this toy model, more research is needed in order to understand the origin, the properties and the value of the creation rate $\Gamma$. 

Going a step forward and without assuming a specific form of $\Gamma$, we can say more about this NMC model by analysing the deceleration parameter defined by $q \coloneqq - \dot H/H^2 - 1$. Thus, using Eq.~\eqref{friedmanneq1} the deceleration parameter for this model is given by
\begin{equation} 
    q =  \frac{\Omega_{M+}}{2}  + \frac{|\Omega_{M-}|}{2}\left[ \left(\frac{\Gamma}{H} -3\right)(1 + w_-) + 2 \right]  - \Omega_\Lambda,  \label{eq:deceleration}
\end{equation}
which is independent of the curvature parameter $k$. Current observations \citep{Riess:1998cb,Perlmutter:1998np} support that the Universe goes through a late cosmic acceleration expansion, so we must have the correct combination of variables in Eq.~\eqref{eq:deceleration} to get $q(z=0) < 0$. For the NMC toy model, with $\Lambda = 0$ and $\Gamma = 3H$, Eq.~\eqref{eq:deceleration} gives 
\begin{equation}
    q =  \frac{1}{2} \Omega_{M+} +  |\Omega_{M-}|,
\end{equation}
which is independent of $w_-$. We notice in here that $q > 0$ is preserved, therefore \textit{the NMC toy model does not allow an accelerated expansion}. This result leads us to reject this toy model as a viable cosmological scenario, since several observations apart from supernovae Type Ia as baryon acoustic oscillations \citep{Bassett:2009mm}, clusters of galaxies \citep{Astier_2012} and gravitational waves as standard sirens \citep{Abbott:2017xzu} along with the standard cosmography approach at low redshifts \citep{Escamilla-Rivera:2019aol} confirm this cosmic acceleration.

Moreover, it is always possible to choose different values for $\Gamma$, $\Omega_\Lambda$ and even $w_-$ (if we try to give a more exotic nature to the negative masses), to get the required accelerated expansion. Nevertheless, the choices for these values might be non-physical and they would significantly increase the complexity of the problem and would erase the initial motivation of a NMC model as a simple alternative explanation of the dark sector.

Following the above general ideas, a final comment can be made for a negative-mass dominated cosmology (i.e. $|\rho_-| \gg \rho_+$) with $\Lambda=0$. From Eq.~\eqref{eq:omegas}, these assumptions imply $k=-1$ and thus it is not always possible to have physical solutions compatible with observations (see comments below), since $H^2$ could take negative values. In particular, in a negative dust Universe without creation of particles ($w_-=0$, $\Gamma=0$), $H^2$ changes sign in 
\begin{equation} \label{eq:a critic}
    a = \frac{|\Omega_{-,0}|}{\Omega_{k,0}},
\end{equation}
with $\Omega_{i,0}=\Omega_i|_{t=0}$, where the subindex $i$ denotes each fluid component. The same pathology occurs in the case $\Lambda >0$, where once again $H^2$ could become negative for a value of $a>0$. By similar arguments, $H^2$ could still be negative when considering radiation and  the creation of negative mass particles, because $\rho_- < 0$ independently of $w_-$ and $\Gamma$, as can be shown from Eq.~\eqref{eq:continuity}. Therefore, within the NMC model we have a Universe which could begin with a finite size. Nevertheless, we know from estimations of the early Universe in the standard $\Lambda$CDM model, that the Universe must begin from a very compact and dense state \citep{weinberg2008cosmology,ellis2012relativistic} to be able to predict cosmological observables, such as the abundances of primordial nuclei \citep{cyburt2016big,steigman2007primordial,iocco2009primordial} and other early universe observations \citep{aghanim2020planck,cruz2021late}. In this way, we must fine-tune the nature of $\rho_-$ and $\Gamma$ in order to make the NMC scale factor compatible with these observables.\\


\textit{Runaway motion}.-
\label{sec:runaway motion}
Considering comoving coordinates and a smooth one-parameter family of geodesics $\gamma_s(t)$, we denote the unit timelike tangent vector field to the family of geodesics as $\mathcal{T}^{a}=(\partial /\partial t)^a$, such that $\mathcal{T}^a\mathcal{T}_a=-1$, and the separation vector between nearby geodesics as $X^a=(\partial /\partial r)^a$, which satisfies the orthogonality property $\mathcal{T}_aX^a=0$. In this way, the geodesic deviation equation is
\begin{equation}\label{eq:geodev}
	A^a=-\tensor{R}{_c _b _d^a}X^b \mathcal{T}^c \mathcal{T}^d,
\end{equation}
where $A^a=\mathcal{T}^c\nabla_c v^a=\mathcal{T}^c\nabla_c \left( \mathcal{T}^b\nabla_b X^a\right)$ is the relative acceleration between nearby geodesics and $\tensor{R}{^a_b _c _d}$ is the Riemann tensor. In particular, it is well known that for fundamental observers in RW space--times, $A^a =-qH^2 (\partial /\partial r)^a$. Since $q=q(t)$ and $H=H(t)$, the relative acceleration must be common to all observers, and due to the symmetries of the Robertson--Walker metric we must deal with other geometries in order to estimate the effects of runaway motion. 

We now consider the particular case of negative and positive particles in an arbitrary space--time. Using the expansion of the Riemann tensor 
\begin{equation}\label{Riemm}
    R_{abcd}=C_{abcd}+ \left(g_{a[c}R_{d]b}-g_{b[c}R_{d]a}\right)-\frac13 Rg_{a[c}g_{d]b},
\end{equation}
where $C_{abcd}$ is the Weyl tensor, and $R_{ab}$ is the Ricci tensor we can rewrite Eq.~\eqref{eq:geodev} as
\begin{eqnarray}
A^a&=&-\left[C_{cbde}+\left(g_{c[d}R_{e]b}-g_{b[d}R_{e]c}\right)-\frac13 Rg_{c[d}g_{e]b}\right]\nonumber\\
&&g^{ea}X^b \mathcal{T}^c \mathcal{T}^d.
\end{eqnarray}
Since the Weyl tensor is related to free space and the Ricci scalar is related to the energy--momentum tensor through the Einstein field equations, we can divide the geodesic deviation equation into an acceleration due to free space $A^a_{\text{FS}}$, and the acceleration due to matter $A^a_{\text{M}}$, defined by
\begin{eqnarray}
A^a_{\text{FS}}& \coloneqq &-\tensor{C}{_c_b_d^a}X^b \mathcal{T}^c \mathcal{T}^d,\\
A^a_{\text{M}}& \coloneqq &-\left[\left(g_{c[d}R_{e]b}-g_{b[d}R_{e]c}\right)-\frac13 Rg_{c[d}g_{e]b}\right] \nonumber \\
& & \, g^{ea}X^b \mathcal{T}^c \mathcal{T}^d. \label{eq:acceleration matter}
\end{eqnarray}
If we consider that the acceleration is due solely to matter and use the Einstein field equations  $R_{ab}=\kappa ( T_{ab}- \frac{1}{2} T g_{ab})$ and $R = - \kappa T$, then Eq.~\eqref{eq:acceleration matter} reads as
\begin{equation} \label{eq:aceleration}
     A^a_{\text{M}} = \frac\kappa2 \left[ \tensor{T}{^a_b}X^b - \mathcal{Q}_b X^b \mathcal{T}^a - \left(\frac{\rho}{3} + 2 p\right) X^a  \right],
\end{equation}
where we have used that $T = g^{ab} T_{ab} = - \rho + 3p$, with the energy density $\rho=u^a u^b T_{ab}$, the isotropic pressure $p=(g^{ab}+u^au^b)T_{ab}$ and the definition of the spatial energy flux vector $\mathcal{Q}_a \coloneqq -\mathcal{T}^b T_{ab}$.

So far, the matter content is completely general. We will restrict to a special case. We consider the matter content as two compact fluids, the first composed of matter with positive density and the second with negative density. These fluids are separated such that we can neglect gravitational effects due to the opposite fluid and thus we can consider both as comoving. With these assumptions, we can rewrite the energy-momentum tensor as  
\begin{equation}\label{eq:tensor rho+ y -}
T_{a b}=(\rho_{+} -|\rho_{-}|)\mathcal{T}_a\mathcal{T}_b,
\end{equation}
and from Eq.~\eqref{eq:aceleration}, we get
\begin{equation}\label{eq:Acc1}
    A^a_\text{M}=\frac{\kappa}{6} \left[-\rho_{+} + |\rho_{-}|\right]  X^a.
\end{equation}
These results can be corroborated with \cite{ellis1999deviation} for positive masses and a FRW geometry. Thus, observers nearby to the positive particles will be attracted while observers in a neighbourhood of the negative particles will be repelled. Physically, this represents the so-called runaway motion. Moreover, using the NMC's effective energy--momentum tensor in Eq.~\eqref{eq:energymom nmc}, $A^a_\text{M}$ reads as
\begin{equation}\label{eq:Acc2}
    A^a_{\text{M}} = \frac{\kappa}{6} \left\{-\rho_+ + |\rho_-| \left[ \frac{\Gamma}{H}(1 + w_-) + 3 w_- +1 \right] \right\} X^a,
\end{equation}
and considering $w_- = 0$ the latter equation can be reduced to
\begin{equation}\label{eq:Acc3}
    A^a_\text{M} = \frac{\kappa}{6} \left[-\rho_+ + |\rho_-| \left( \frac{\Gamma}{H} +1 \right) \right] X^a,
\end{equation}
therefore, we obtain analogous results to the dust case. We notice that, unlike the fundamental observers in RW space--times, the  relative acceleration in Eqs.~\eqref{eq:Acc1}-\eqref{eq:Acc3}, is not homogeneous nor isotropic, hence in a NMC model with $|\rho_-|\left( \frac{\Gamma}{H} +1 \right) > |\rho_+|$, we have \textit{a positive relative acceleration between nearby geodesics, which could cause non-linear effects in the velocity perturbations. Thus, structure formation in this kind of cosmology becomes a fine-tuning problem.}
\\


\textit{Concluding remarks}.-\label{sec:conclusions}
In this letter we thoroughly analysed a negative mass  cosmological (NMC) model which arise as an alternative, simpler and unified explanation of the dark sector of the Universe, within the context of General Relativity, and with a possible background of continuously created negative mass particles. We found that this model requires fine-tuning in order to predict a late cosmological accelerated expansion, to have a Hubble parameter defined in the whole cosmic redshift range expected from early Universe observations, and to allow structure formation. Moreover, we discarded the reviewed model with a negative mass creation rate of $\Gamma = 3H$, because it does not predict late accelerated expansion; this particular choice of $\Gamma$ had been used to explain and numerically model the effects of dark energy and dark matter. In its current state, the NMC model requires a more concise theoretical formulation before being considered as a viable model of the Universe which can be tested with cosmological surveys.

\rule{70mm}{0.6mm}

\textit{Acknowledgments}.-
AA, AG and SN acknowledge financial support from CONACYT postgraduate grants program.
CE-R acknowledges the Royal Astronomical Society as FRAS 10147 and by DGAPA-PAPIIT-UNAM Project IA100220. This article is also based upon work from COST action CA18108, supported by COST (European Cooperation in Science and Technology). The ideas treated in this Letter were derived from a discussion in the Lecture entitled \textit{``Aplicaciones Astrof\'isicas y Cosmol\'ogicas  de la Relatividad General''} at ICN-UNAM.


\bibliographystyle{aa}
\bibliography{references}
\end{document}